\documentclass[letter]{aa} 
\usepackage{txfonts}
\usepackage{url}
\usepackage{natbib}
\bibpunct{(}{)}{;}{a}{}{,} 
\usepackage[utf8]{inputenc}
\usepackage{subfigure}
\usepackage{verbatim} 
\begin{document}

   \title{The relation between photospheric supergranular flows and magnetic flux emergence}
   \author{M. Stangalini}
   \institute{INAF-Osservatorio Astronomico di Roma, 00040 Monte Porzio Catone (RM), Italy\\
   \email{marco.stangalini@inaf.it}}

  \abstract 
{A recent study carried out on high sensitivity SUNRISE/IMAX data has reported about the existence of areas of limited flux emergence in the quiet Sun. By exploiting an independent and longer ($4$ hours) data set acquired by HINODE/SOT,  we further investigate these regions by analysing their spatial distribution and relation with the supergranular flow. Our findings, while confirming the presence of these calm areas, also show that the rate of emergence of small magnetic elements is largely suppressed at the locations where the divergence of the supergranular plasma flows is positive. This means that the dead calm areas previously reported in literature are not randomly distributed over the solar photosphere but they are linked to the supergranular cells themselves. These results are discussed in the framework of the recent literature.}  

   \authorrunning{M. Stangalini}
	\titlerunning{Supergranular flows and flux emergence}
\maketitle

\section{Introduction}
Recently, \citet{2012ApJ...755..175M} have identified in the quiet photosphere areas of reduced flux emergence on spatial scales larger than granules ($\sim 1000$  km), which they named "dead calm areas". These areas also show a moderate deficit of circular polarization signals in the deep magnetograms obtained from the same data by \citet{2013SSRv..tmp...21M}.\\
It has been found that small magnetic elements with diameters comparable to or below the present resolving power of the current solar telescopes ($ \sim 100-150$ km on the solar surface) are abundant in the very quiet Sun areas of the solar photosphere \citep{2007ApJ...670L..61O, 2008ApJ...672.1237L, 2012ApJ...747L..36V, 2012ApJ...758L..40M}. 
   \begin{figure*}[!t]
   \centering
   \subfigure[Horizontal velocity field] {\includegraphics[width=8.cm, clip]{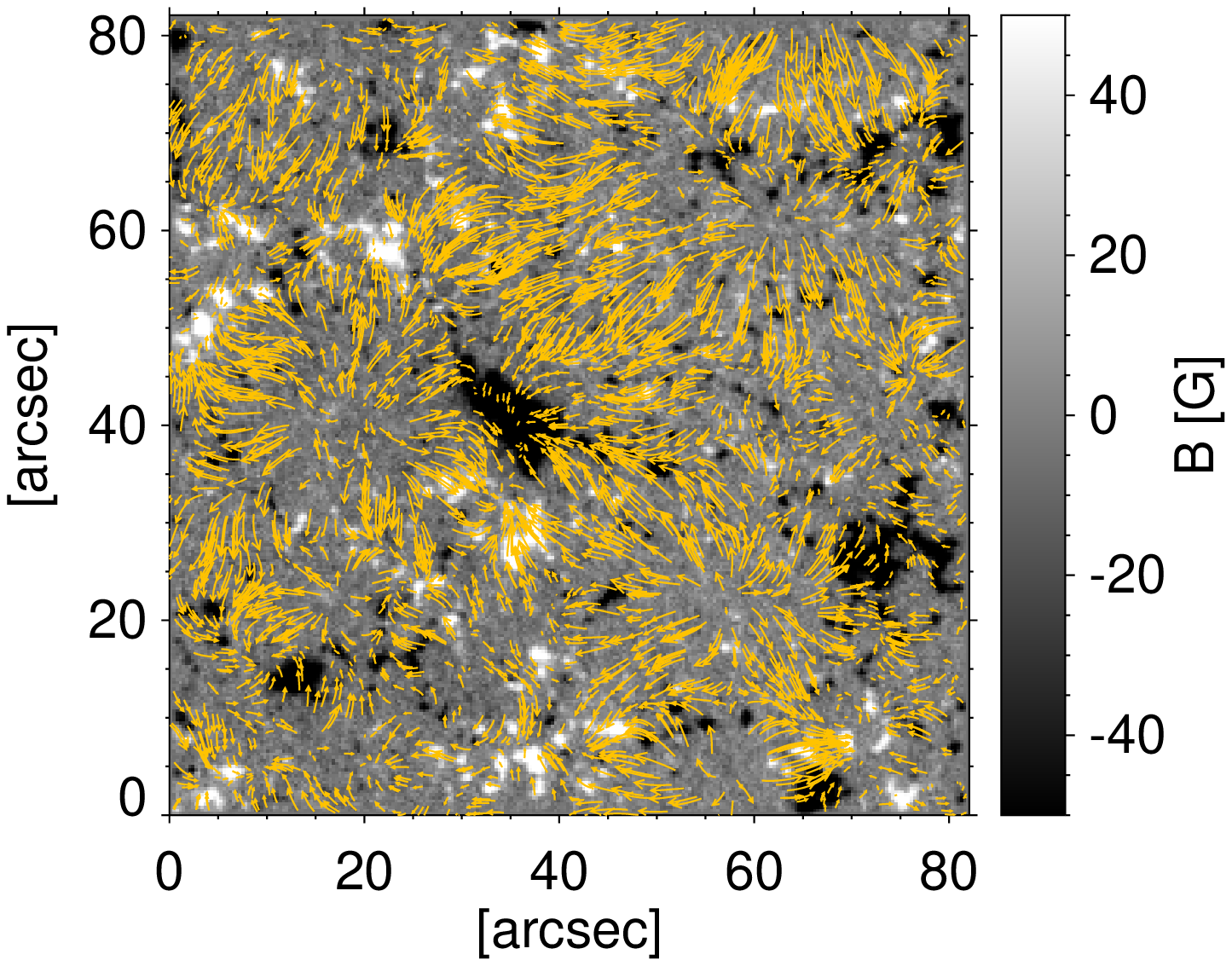}}
   \subfigure[Map of divergence and flux emergence] {\includegraphics[width=8.cm, clip]{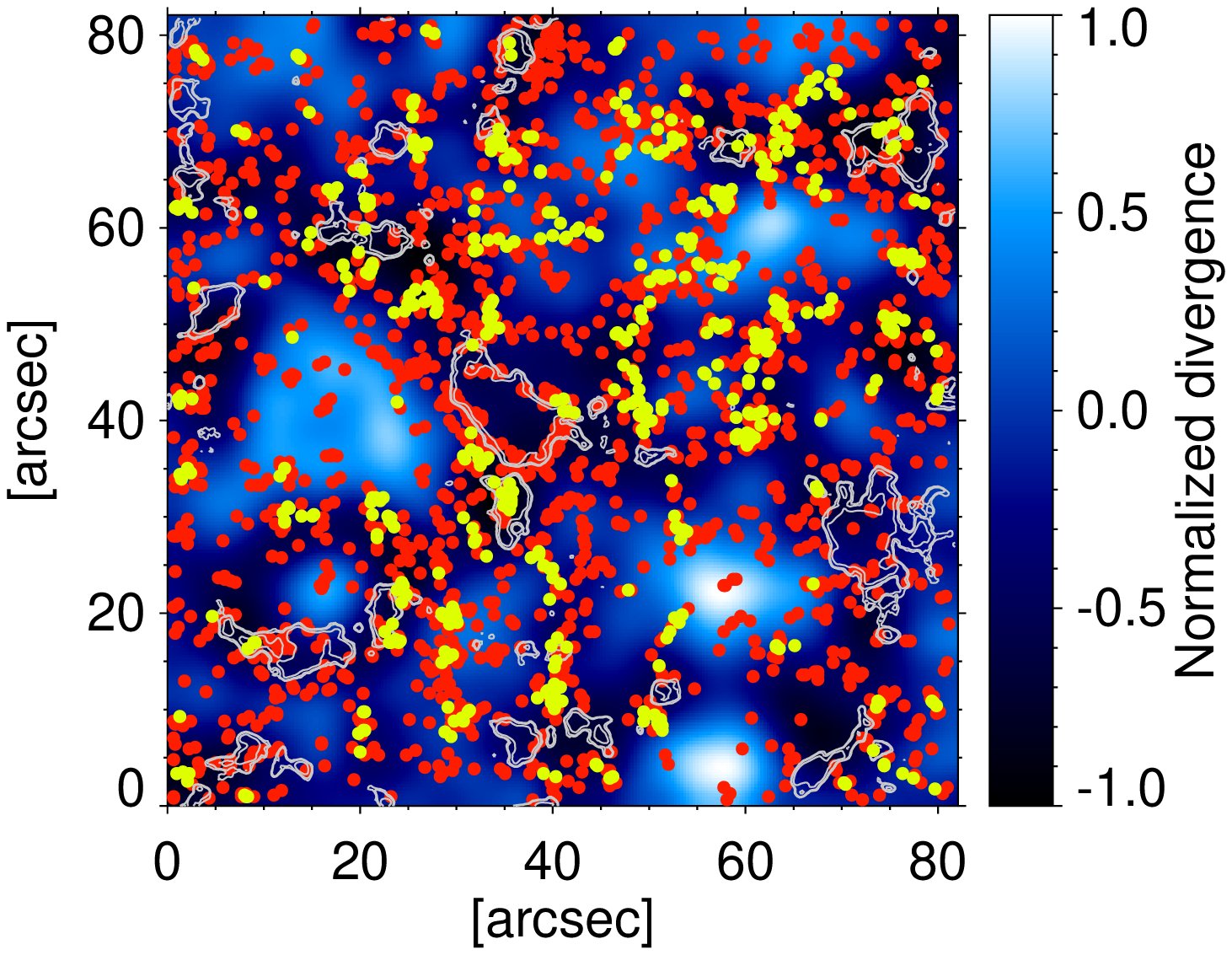}}
   \subfigure[Deep magnetogram] {\includegraphics[width=8.cm, clip]{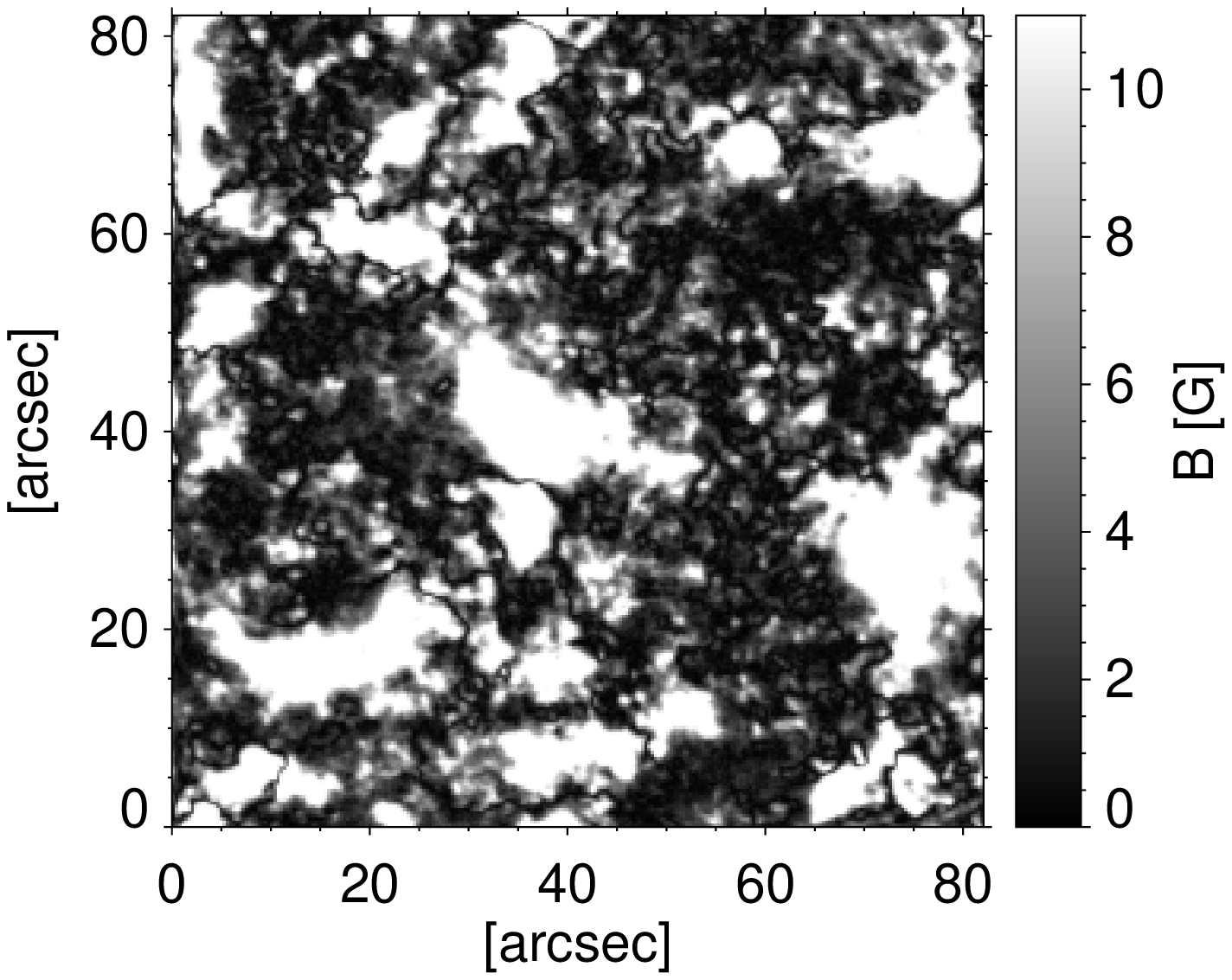}}
   \caption{(Panel a) Time averaged magnetogram obtained by averaging over the total duration of the data set with superimposed the horizontal velocity field obtained from the local correlation tracking (LCT, orange arrows). (Panel b) Normalized divergence map obtained from LCT with the location of magnetic appearance events over the whole duration of the data set (red dots), and flux emergence events (yellow dots). The white contours indicate the location of the maxima of the absolute value of the  magnetic field ($|B|>50$ G). (Panel c) Deep absolute magnetogram saturated at $11$ G. } 
    \label{map}
   \end{figure*}    
They have been observed to emerge in small magnetic loops within supergranules \citep[e.g.][]{2007ApJ...666L.137C,2009ApJ...700.1391M} and then diffuse radially toward the edges of the supergranules themselves \citep{2011ApJ...743..133A, 2012ApJ...758L..38O, 2012ApJ...759L..17L, 2013ApJ...770L..36G}, to form the so-called magnetic network \citep{1989ApJ...345.1060S}. In particular, \citet{2012ApJ...758L..38O}, by exploiting a very long series of high resolution magnetograms ($8$ hours), have observed the small magnetic concentrations within a supergranule moving radially outward from the center of the supergranule toward the network, with a velocity aligned to the plasma flow. Their results have also showed that the dynamical properties of these magnetic concentrations depend on their location within the supergranule itself. They indeed accelerate and then decelerate in proximity of the supergranular boundaries. The same author have argued that this may affect the spatial distribution of the magnetic flux observed, as a direct consequence of magnetic diffusion being more effective in some areas of the supergranular cells. To this regard, \citet{2013ApJ...770L..36G} have observed a change of the diffusion regime of small magnetic elements with a reduction of the diffusivity at spatial scales larger than granulation and comparable with supergranular scales. \\ 
Small magnetic elements in the solar photosphere are also though to be of some importance in deciphering different unsolved question in  solar physics, as for example the heating of the upper layers of the Sun's atmosphere through the excitation of MHD waves \citep[see for example][]{2011ApJ...730L..37M, 2013A&A...549A.116J, 2013A&A...554A.115S, 2013A&A...559A..88S} and their contribution to the solar irradiance \citep{2012ASPC..463...65R}. 
Although as seen before they have been studied in detail in many aspects, their origin is still a matter of debate. 
Indeed, two main competing theories have been put forward to explain the presence of this small scale magnetic fields in the solar photosphere. In the first scenario these magnetic fields are the end-product of the decaying of larger active regions produced by the global dynamo. In the second scenario the magnetic elements are generated by a local small scale dynamo (SSD) driven by granular and supergranular plasma flows. Actually both mechanisms are likely to contribute at the same time, although it has to be determined which is the dominant one \citep{Cattaneo01062001}.\\
However, in the last few years, many observational results have been put forward in favor to the presence of a small-scale dynamo \citep[e.g.][]{2008A&A...481L..25I, 2010ApJ...723L.149D, 2011ApJ...737...52L, 2013A&A...555A..33B}, as well as results from numerical simulations of the solar convective zone \citep{2007AA...465L..43V, 2008A&A...481L...5S, 2010ApJ...714.1606P, 2010A&A...513A...1D}. For a deeper examination of the observational evidences in favour to the SSD we refer the reader to the review by \citet{2013SSRv..tmp...21M}. It is unclear, however, how the existence of such a calm areas observed by \citet{2012ApJ...755..175M} could fit this general framework, although they represent an important constraint on the generation and amplification of the magnetic field on small scales in the solar photosphere. Among the many possible explanations for the existence of these areas, one interesting possibility, indicated by the authors themselves, is that this could be due to the highly intermittent character of the SSD \citep{Cattaneo01062001}, which leads to a bursty appearance of small-scale magnetic fields. Moreover, \citet{2008PhRvL.101l5003T} have demonstrated that coherent flows with their sustained stretching are far more efficient, in terms of dynamo action, than incoherent plasma flows.  For this reason the same authors argued that the role of coherent structures should be always carefully assessed when dealing with small scale dynamos. \\
The IMAX data analysed by \citet{2012ApJ...755..175M} had a limited temporal duration ($\sim 30$ min). This prevented any possible analysis of the calm areas in relation to large scale (tens of Mm) plasma flows such as those determining the supergranular patterns. 
In this study we take advantage of the long temporal duration of the data provided by SOT/FG, the narrow band imager on board Hinode satellite \citep{springerlink:10.1007/s11207-008-9174-z}, to further investigate the existence of these areas and to assess the role of supergranular flows on their origin. 
   
\section{Data set and analysis}
The data set used in this study consists of a sequence of high spatial resolution magnetograms and intensity measurements acquired by SOT/NFI in the Na I $589.6$ nm spectral line that is sampled at two wavelength positions from the line center ($\pm 160 ~m\AA$). The magnetograms were generated from shutterless V and I Stokes filtergrams taken on 2008 August 18, close to the disc centre. The temporal cadence of the data is $30$ s. The data are slightly downsampled since the pixel scale is set at $0.16$ arcsec, while the diffraction limit is $\sim 0.24$ arcsec. The field of view (FoV) is approximately $80 \times 80$ arcsec, and the total duration of the time sequence is approximately $4$ hours. In Fig. \ref{map} (panel a) the time averaged magnetogram saturated between $-50$ G and $50$ G is shown. The FoV encompasses a few supergranules whose boundaries are highlighted by the network patches visible in the figure.\\
In addition to the standard calibration procedure  (the IDL code \textit{fg-prep.pro} available in the Hinode Solarsoft package), a registration procedure allowing subpixel accuracy was applied. This was done to account for any possible residual misalignment due to satellite's jitter and tracking errors. This procedure is based on FFT cross-correlation and utilizes the whole FoV to estimate the misalignment between two images. We applied iteratively the FFT registration, until the mean residuals were minimized (until a stable value was reached). This happened in three iterations.\\
To study the emergence of small magnetic features we employed the YAFTA tracking algorithm \citep{Welsch2003, 2007ApJ...666..576D} which allows us to identify and track magnetic pixels belonging to the same local maximum. For tracking purposes two constraints were applied. Only features lying above a threshold in the magnetograms and having an area as small as the sampling resolution limit set by the pixel scale were considered ($2 \times 2$ pixels in our case corresponding to $240 \times 240$ km on the solar photosphere). The threshold on the magnetic signal is chosen to be $2 \sigma$ (equivalent to $11.8$ G). Following \citet{2012SoPh..279..295L} we estimated the sigma of the magnetic signal by fitting a Gaussian to the low-field pixels with absolute value of the magnetic flux density below $200$ G. 
With YAFTA tracking code, we identified the newly appeared magnetic elements and studied their initial position in the field of view. Among the complete sample of newly appeared magnetic elements fulfilling the above criteria, we only retained those lasting (trackable) at least three temporal steps ($\sim 90$ s). This is done to avoid spurious detections. \citet{2010ApJ...720.1405L} have shown that up to of $75 \%$ of the newly appeared magnetic elements in magnetograms can be ascribed to coalescence, in which a magnetic flux tube undergoes a field intensification, thus becoming detectable. This makes the detection of emergence events difficult. Since the tracking algorithm does not discriminate between different physical processes leading to the appearance of new magnetic features we used the Tracked Bipolar and Cluster Method (hereafter TBCM) developed and used in \citet{2005SoPh..231...45C, 2005GApFD..99..513C, 2011SoPh..269...13T} to identify emergences. This method, starting from the tracked elements and using the birth information provided by the tracking code, identifies newly emerged bipoles and clusters and selects unique pairings among them. More in detail, the code looks for features that satisfy the following conditions: the distance between two opposite polarity features $D_{0}$  and the temporal difference of appearance $\Delta T$ must be smaller than a fixed quantity, and the ratio of the fluxes $\rho$ must satisfy the relation
$1/ \rho \leq |\phi_{i+}|/|\phi_{i-}| \leq \rho$, where $\phi_{i+}$ and $\phi_{i-}$ represent the magnetic fluxes associated to the footpoints of the bipole. At this point, not all the identified emerging bipoles and clusters with the TBCM method represent a unique feature pairings (emerging loops or biboles). This can be accomplished by using a connectivity matrix as in \citet{2005GApFD..99..513C, 2011SoPh..269...13T}. Similar to \citet{2011SoPh..269...13T} we chose $\rho=3$, $\Delta T=15$ min, and $D_{0}=1.3$ Mm. Our aim is the assessment of possible effects of supergranular flows on the spatial distribution of flux emergence.
For this purpose we have estimated the supergranular flow field by using a local correlation tracking technique (LCT). In particular we made use of the FLCT code \citep[Fast Local Correlation Tracking, ][]{2008ASPC..383..373F} with a spatial window of $12$ pixels, or equivalently, $\simeq 1400$ km that is a scale slightly larger than the granular scale. The granules are in fact though to be a good tracers of the large scale flows. Each intensity image of the sequence is used in the LCT, thus the original temporal cadence is maintained. Finally the horizontal velocity fields at each time step are then used to estimate the average velocity field over the whole duration of the data set ($\sim 4$ hours). The final horizontal velocity field is additionally smoothed with a running window of $10 \times 10$ pixels ($1200 \times 1200$ km) to further reduce the granular noise. The final result of this analysis is overplotted in Fig. \ref{map} (panel a, yellow arrows). The FoV encompasses a few supergranules as clearly indicated by the direction of the horizontal flows.

\section{Results}
\begin{figure}[!ht]
   \centering
   \subfigure{\includegraphics[width=9cm, clip]{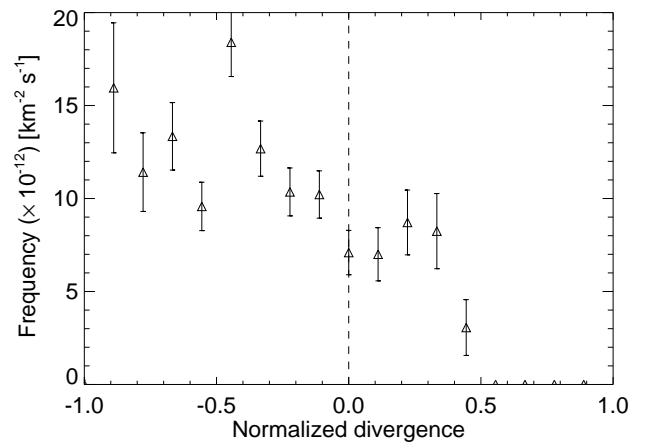}}
   \caption{Frequency of flux emergence derived from tracking as a function of the divergence of the flow field estimated from the LCT. The error bars represent the standard error of the mean. } 
    \label{plot}
   \end{figure}

We studied the spatial distribution of emergence events, which was derived using the information provided by the tracking algorithm together with the TBCM method described above, looking for possible relation with the horizontal velocity fields obtained from the LCT.
Fig. \ref{map}, panel b, shows the position of the newly appeared magnetic elements (red dots), and disambiguated newly emerged bipoles and clusters (yellow dots) obtained from the TBCM code, overplotted on the divergence map obtained from LCT.  In agreement with \citet{2012ApJ...755..175M}, we found that the magnetic elements emerge non-homogeneously giving raise to areas lacking of frequent flux emergence.  Besides, the comparison with the divergence map shows that these areas occupy regions characterized by positive divergence (sources of the supergranular flow field). Panel c of the same figure also depicts the mean deep magnetogram (saturated at $\pm 11$ G) to be compared with the divergence map of panel b. Similar to what was reported by \citet{2013SSRv..tmp...21M}, the deep magnetogram also shows a lack of residual magnetic signal at the same locations at which a deficit of emergence is seen. Even in this case, it is straightforward to see how these regions correspond to areas with positive divergence. This can be better seen in  Fig. \ref{plot} where we plot the frequency of emergence of the magnetic elements selected with the TBCM method as a function of the divergence. The frequency of emergence is smaller at the locations of positive divergence (centers of supergranules) and larger at the boundaries of the convective cells (negative divergence).

\section{Discussions and conclusions}
In this work by analysing a long series of high cadence and high spatial resolution magnetograms acquired with SOT/FG, we have studied the spatial distribution of flux emergence in supergranules. In particular we have investigated the frequency flux emergence in relation to the supergranular flows. Very long time series are indeed required to disentangle the supergranular from granular flows in LCT analyses. From newly appeared magnetic elements provided the YAFTA tracking code, we have selected only the flux emergence cases  by means of the TBCM method. As underlined in \citet{2011SoPh..269...13T}, this method relies on strict criteria for selecting emerging bipoles and clusters of magnetic elements.
Our results confirm the presence of the surprisingly quieter magnetic areas in the solar photosphere, which were previously reported by \citet{2012ApJ...755..175M} in SUNRISE/IMAX data. Our results are also consistent with the findings by \citet{1988SoPh..117..343W} who observed a slight tendency of the magnetic fields to emerge close to the supergranular boundaries. 
In addition to the results of \citet{2012ApJ...755..175M} we have also found that these dead calm areas coincide with locations of positive divergence of the supergranular plasma flows. We have found that the flux emergence rate is indeed maximally reduced at the centers of supergranules while it increases toward the borders of them. This behaviour is also seen in deep magnetograms where a lack of residual magnetic signal can be found in correspondence with the centers of supergranules. Although this deficit of residual magnetic signal was already reported by \cite{2013SSRv..tmp...21M},  no link with supergranular flows was possible in that case due to the limited duration of the data sets analysed.\\
Framing these results in a SSD scenario is difficult \citep{2013SSRv..tmp...21M}, however it has been shown that the presence of coherent flows and velocity patterns leads to the development of efficient dynamos in comparison with random flows. Indeed, \citet{2008PhRvL.101l5003T} have demonstrated that the amplification of the magnetic field in a SSD is strongly dependent on the presence of spatially and temporally coherent structures.  In particular when two velocity flows with the same spectra are considered, the one with larger spatially and temporally coherent structures result in a more efficient SSD mechanism. In addition, the same authors have shown that the presence of coherent velocity flows leads to the generation of less homogeneous magnetic field distributions owning to the line stretching of the coherent flows themselves.\\
Although one is tempted to believe the granular flows as the most obvious patters giving birth to the small scale magnetic fields detected in high resolution and high sensitivity observations, the supergranulation has been also indicated as a possible mechanism from which a dynamo action may originate \citep{1999ApJ...515L..39C}.  This, together with the above scenario of enhanced amplification produced by regular flows may constitute an important ingredient to be considered for explaining our observational findings. \\
\citet{2013SSRv..tmp...21M} has already pointed out that the voids of flux emergence observed by \citet{2012ApJ...755..175M} are barely consistent with a granularly driven SSD since they are over larger scales ($> 1$ Mm). To this purpose it is worth noting that \citet{2012ApJ...745..160G} have studied a magnetic bipole emerging in the solar photosphere and reported a footpoint separation of $\sim 4$ Mm which is also larger than the granular scale, thus inconsistent with a SSD driven by granular convection only. \\
Our results also fit this picture by demonstrating a strong non-homogeneous emergence pattern of the magnetic field over the FoV. This is not surprising in light of the results of \citet{2008PhRvL.101l5003T} that demonstrated that the process of stretching and amplification of the field through an SSD can be highly non-local and patchy depending on the properties of the coherent structures of the velocity field.  Due to the finite sensitivity of the instrumentation, we cannot argue that at the center of supergranules the flux emergence is totally inhibited. It can well be that the emerging magnetic field generated by an SSD is far from being detectable at that locations, probably because of a less efficient amplification owing to the supergranular pattern and its dynamics.  However, our results suggest a relation between the flux emergence rate (and or the amplification of the field) and the supergranular flows. \\
To this regard, another and perhaps more intriguing explanation for our findings may come from different considerations. 
As mentioned above, \citet{2012ApJ...758L..38O} have argued that since the dynamical properties (i.e. the radial velocity) of magnetic elements depend on their position within the supergranule, the spatial distribution of magnetic fields observed might be affected by this resulting in an inhomogeneous distribution. This is a direct consequence of the magnetic diffusion being more effective in certain locations. 

\begin{acknowledgements}
This work has been supported by the PRIN-INAF 2010 grant, funded by the Italian National Institute for Astrophysics (INAF). We thank Francesco Berrilli and Fabio Giannattasio for useful discussions. We also thank Ilaria Ermolli for a critical reading of the first draft of this manuscript and the anonymous referee for providing useful comments and suggestions which helped a lot to improve the first version of the paper. Hinode is a Japanese mission developed and launched by ISAS/JAXA, collaborating with NAOJ as a domestic partner, NASA and STFC (UK) as international partners. Scientific operation of the Hinode mission is conducted by the Hinode science team organized at ISAS/JAXA. This team mainly consists of scientists from institutes in the partner countries. Support for the post-launch operation is provided by JAXA and NAOJ (Japan), STFC (U.K.), NASA, ESA, and NSC (Norway).\\
\end{acknowledgements}

\bibliography{lib}
\end{document}